\documentclass{aa}  
\usepackage{natbib}
\usepackage{graphicx}
\usepackage{txfonts}
\newcommand{\referee}{\rm}

\begin{document}
\title{Opacity in compact extragalactic radio sources and its effect on
astrophysical and astrometric studies}
\titlerunning{Opacity in compact extragalactic radio sources}
\author{Y.~Y.~Kovalev\inst{1,2}
       \and
       A.~P.~Lobanov\inst{1}
       \and
       A.~B.~Pushkarev\inst{1,3,4}
       \and
       J.~A.~Zensus\inst{1}
       }
\authorrunning{Kovalev et~al.}
\offprints{Y.~Y.~Kovalev}
\institute{Max-Planck-Institut f\"ur Radioastronomie,
           Auf dem H\"ugel 69, 53121 Bonn, Germany\\
           \email{ykovalev,
                  alobanov,
                  apushkar,
                  azensus@mpifr-bonn.mpg.de}
           \and
           Astro Space Center of Lebedev Physical Institute,
           Profsoyuznaya 84/32, 117997 Moscow, Russia
           \and
           Central (Pulkovo) Astronomical Observatory,
           Pulkovskoe Chaussee 65-1, St.~Petersburg 196140, Russia
           \and
           Crimean Astrophysical Observatory, Crimea, Ukraine
           }

\date{Received September~17,~2007 / Accepted February~16,~2008}

\abstract
{
The apparent position of the ``core'' in a parsec-scale radio jet (a
compact, bright emitting region at the narrow end of the jet) depends on
the observing frequency, owing to synchrotron self-absorption and
external absorption. While providing a tool probing physical conditions
in the vicinity of the core, this dependency poses problems for
astrometric studies using compact radio sources.
}
{
We investigated the frequency-dependent shift in the positions of the
cores ({\em core shift}) observed with very long baseline interferometry
(VLBI) in parsec-scale jets.  We discuss related physics, as well as its
effect on radio astrometry and the connection between radio and optical
positions of astrometric reference objects.
}
{
We searched for the core shift in a sample of 277 radio sources
imaged at 2.3~GHz (13~cm) and 8.6~GHz (4~cm) frequency bands using VLBI
observations made in 2002 and 2003. The core shift was measured by
referencing the core position to optically thin jet features whose
positions are not expected to change with frequency. 
}
{
We present here results for 29 selected active galactic nuclei (AGN)
with bright distinct VLBI jet features that can be used in differential
measurements and that allow robust measurements of the shift to be made. In
these AGN, the magnitude of the measured core shift between 2.3 and
8.6~GHz reaches 1.4~mas, with a median value for the sample of 0.44~mas.
Nuclear flares result in temporal variability of the shift.
}
{
An average shift between the radio (4~cm) and optical (6000~\AA) bands
is estimated to be approximately 0.1~mas,
and it should be taken into account in order to
provide the required accuracy of the radio-optical reference frame
connection. This can be accomplished with multi-frequency VLBI
measurements yielding estimates of the core shift in the sources used for
the radio reference frame and radio-optical position alignment.  
}

\keywords{
galaxies: active~---
galaxies: jets~---
radio continuum: galaxies~---
astrometry~---
reference systems
}
\maketitle
%

\section{Introduction \label{s:intro}}

\begin{figure*}[t!]
\begin{center}
\resizebox{0.9\hsize}{!}{
   \includegraphics[trim=0cm 0cm 0cm 2.8cm,clip]{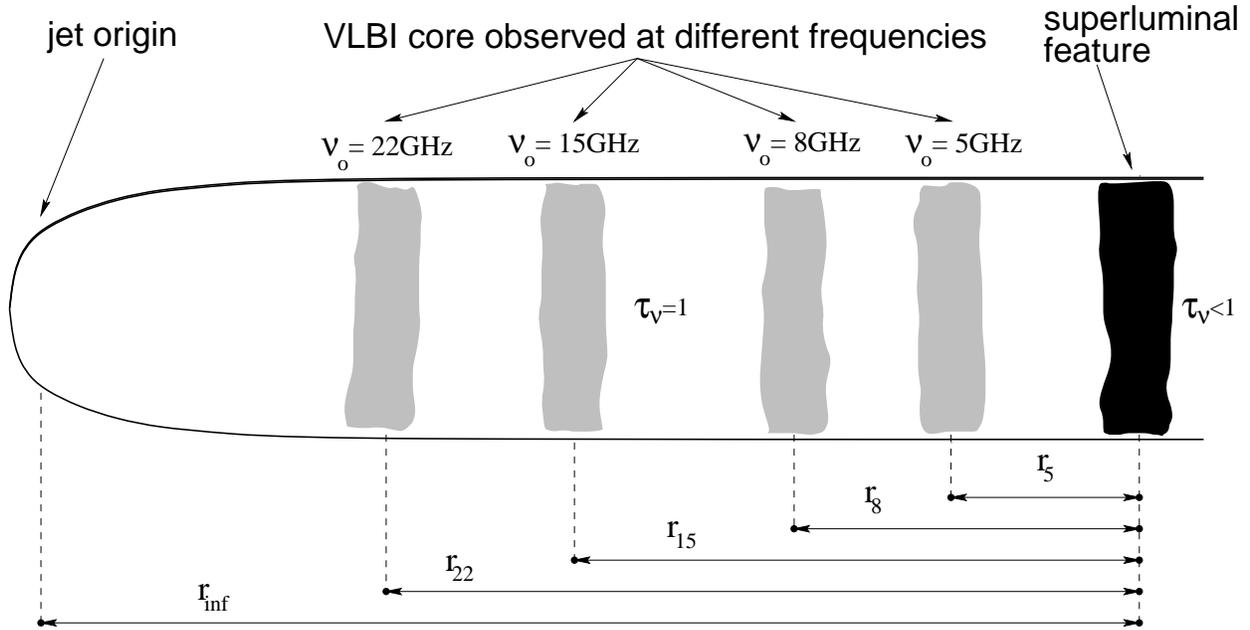}
}
\end{center}
\caption{
\label{f:art}
A scheme illustrating the frequency-dependent position shift
of the VLBI core. Adopted from \cite{L96}.
}
\end{figure*}

Extragalactic relativistic jets are formed in the immediate vicinity of
the central black holes in galaxies, at distances on the order of 100
gravitational radii, and they become visible in the radio at distances
of about 1000 gravitational radii \citep{LZ2007}. This apparent origin
of the radio jets is commonly called the ``core''. In radio images of
extragalactic jets, the core is located in the region with an optical
depth $\tau_s\approx 1$. This causes the absolute position of the core,
$r_\mathrm{core}$, to vary with the observing frequency, $\nu$, since
the optical depth profile along the jet depends on $\nu$:
$r_\mathrm{core} \propto \nu^{-1/k_\mathrm{r}}$
\citep{BlandfordKonigl79}. See an illustration of this effect in
Fig.~\ref{f:art}. Variations in the optical depth along the jet can
result from synchrotron self-absorption \citep{Koenigl81}, pressure and
density gradients in the jet and free-free absorption in the ambient
medium most likely associated with the broad-line region (BLR)
\citep{L98}. If the core is self-absorbed and in equipartition, the
power index $k_\mathrm{r}=1$ \citep{BlandfordKonigl79}. Density and
pressure gradients in the jet and external absorption can lead to
deviations in $k_\mathrm{r}$ from unity \citep{L98}. Changes in the core
position measured between three or more frequencies can be used to
determine critical physical and geometrical parameters of the
relativistic jet origin.

\begin{figure*}[t!]
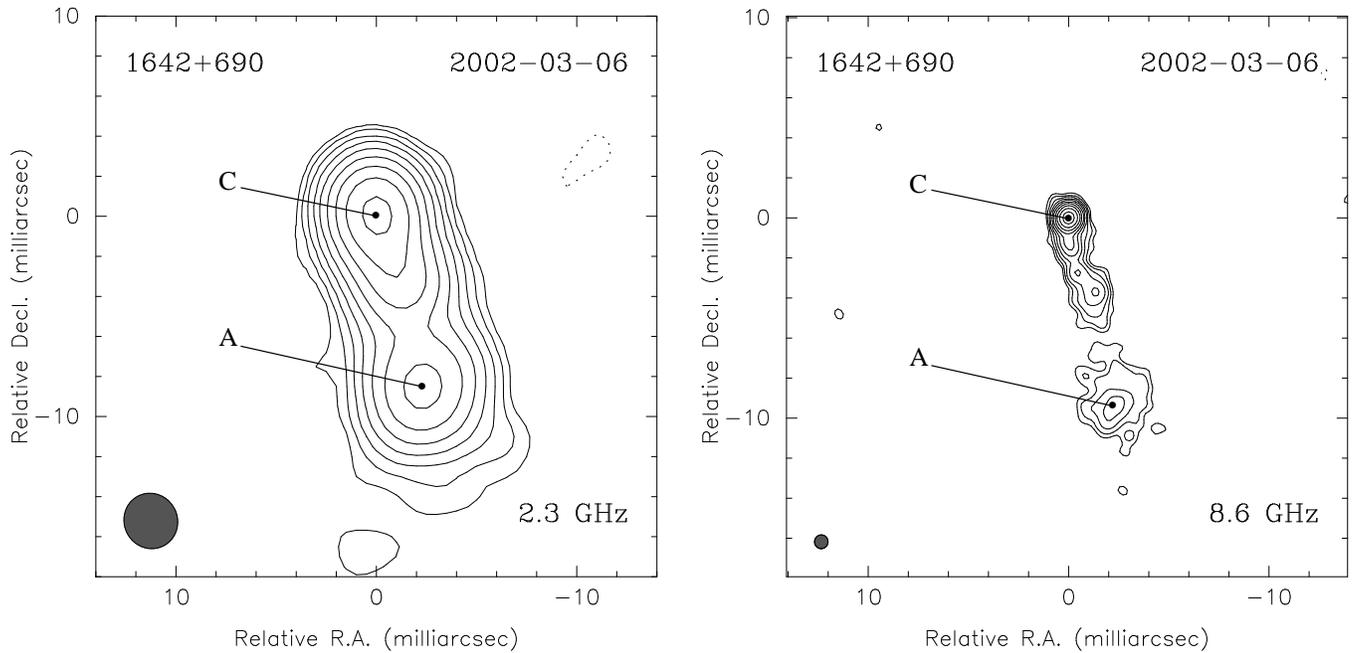

\begin{center}
\resizebox{\hsize}{!}{
   \includegraphics[trim=0.0cm 5cm 0cm 2cm,clip]{J1642+6856_S_2002_03_06_pus_map.ps}
   \includegraphics[trim=0.0cm 5cm 0cm 2cm,clip]{J1642+6856_X_2002_03_06_pus_map.ps}
}
\end{center}
\caption{
\label{f:J1642}
\referee
Example of global VLBI results. 2.3 and 8.6~GHz naturally weighted
images of the quasar J1642+6856, epoch 6~March, 2002. 
The beam, with typical 
full width at half maximum
of 2.6~mas at 2.3~GHz and 0.7~mas at 8.6~GHz, is
shown in the lower left hand corner of each map. The contours are
plotted in successive powers of 2 times the lowest contour.
The peak intensity and the lowest contour for the
2.3~GHz and 8.6~GHz images are
367 and 1.2~mJy\,beam$^{-1}$,
752 and 0.8~mJy\,beam$^{-1}$,
respectively.
The core (`C') position is measured relative to the optically thin jet
feature `A'. The spectral index of the core is
$\alpha=0.6$, while that of the jet feature
$\alpha=-0.6$ ($S\propto\nu^\alpha$).
}
\end{figure*}

The core shift induced by synchrotron self-absorption extends over the
entire spectral range and has an immediate effect on astrometric
measurements made in the radio and optical domains. One particular issue
likely to be affected by the core shift is the connection between the
radio and optical positions of distant quasars. At present, the
International Celestial Reference System (ICRS) is realized primarily by
the International Celestial Reference Frame
\citep[ICRF,][]{icrf98,icrf-ext2-2004} based on very long baseline
interferometry (VLBI) measurements at 2.3\,GHz (13\,cm band) and
8.6\,GHz (3.6\,cm band). The present extension of the ICRF to visible
light is based on the {\em Hipparcos} Catalogue, with rms uncertainties
estimated to be 0.25\,mas\,yr$^{-1}$ in each component of the spin
vector of the frame and 0.6\,mas in the components of the orientation
vector at the catalogue epoch, J1991.25 \citep{Kovalevski_etal97}. The
link between the VLBI and {\em Hipparcos} reference frames and its
accuracy is discussed by \cite{Lestrade_etal95}. In the near future, new
optical astrometric reference frames will be established by the {\em
GAIA} astrometry mission \citep{LP96} and the {\em Space Interferometry
Mission} \citep[{\em{}SIM},][]{SIM98}. Both missions will explore
largely the 3000\,\AA--10000\,\AA\ range corresponding to
(0.3--$1)\times 10^{6}$~GHz. {\em GAIA} will measure optical positions
of about 500,000 distant quasars with $\lesssim 0.1$\,mas accuracy,
similar to the accuracy of radio source positions from the ICRF and the
VLBA\footnote{Very Long Baseline Array of the National Radio Astronomy
Observatory} Calibrator Survey \citep{vcs1,vcs2,vcs3,vcs4,vcs5,vcs6}. {\em
SIM} is expected to deliver high precision astrometric positions of
bright quasars at $\approx 10\,\mu$as accuracy and a target
accuracy of $\approx 20\,\mu$as is envisaged for the radio-optical
alignment \citep{unwin2005}. For both of these missions, an important
problem will be matching the optical catalogues to the radio astrometry
catalogues based on precise positions of compact extragalactic objects and
determining mutual rotations, distortions and zonal systematic errors
\citep[e.g.][]{Fey_etal2001,Souchay_etal06,Lambert_etal06,Frey_etal06,Bourda_etal2007}.
This match relies on an assumption that the dominating component of
emission in both the radio and optical bands is physically the same
region (see sect.~\ref{s:radio-optics} for detailed discussion).

The core shift is expected to introduce systematic offsets between the
radio and optical positions of reference sources, affecting strongly the
accuracy of the radio-optical matching of the astrometric catalogues.
The magnitude of the core shift can exceed the inflated errors of the
radio and optical positional measurements by a large factor. This makes
it necessary to perform systematic studies of the core shift in the
astrometric samples in order to understand and remove the contribution
of the core shift to the errors of the radio-optical position
alignment.

Measurements of the core shift have been done so far only in a small
number of objects
\citep[e.g.,][]{MES94,Lara_etal94,PR97,L96,L98,PFF2000,RL2001,BBR2004,Kadler_etal04}.
In this paper, we present for the first time results for 29 compact
extragalactic radio sources used in VLBI astrometric studies. In
Sect.~\ref{s:res} we describe global VLBI data, the source sample, and
the method of measurement used in this paper, and present results of the
core shift measurements. In Sect.~\ref{s:pa} we discuss astrophysical
applications of the frequency-dependent core shift (Sect.~\ref{s:phys}),
variability of the shift resulting from nuclear flares
(Sect.~\ref{s:variability}), effect of the shift on multi-frequency VLBI
studies (Sect.~\ref{s:multiVLBI}). We investigate the influence of this
effect on the radio-optical reference frame matching and suggest a
method to compensate for it (Sect.~\ref{s:radio-optics}). We summarize
our results in Sect.~\ref{s:sum}.

\section{Core shift measurements \label{s:res}}

We have imaged and analyzed 277 sources from geodetic
RDV\footnote{Research and Development VLBA experiments
\citep[see, e.g.,][]{Gordon05}.} observations made in 2002 and 2003~---
ten 24\,hr-long experiments on
16~January~2002,
6~March~2002,
8~May~2002, 
24~July~2002,
25~September~2002,
11~December~2002,
12~March~2003,
7~May~2003,
18~June~2003.
Geodetic RDV sessions feature simultaneous observations at 2.3~GHz and
8.6~GHz (S and X bands) with a global VLBI network
at right circular polarization.
This includes for
every session the VLBA and up to nine other radio telescopes from the
following list: Algonquin Park (46~m), Gilcreek (26~m), HartRAO (26~m), Kokee
(20~m), Matera (20~m), Medicina (32~m), Noto (32~m), Ny~Alesund (20~m),
Onsala (20~m), TIGO (6~m), Tsukuba (32~m), Westford (18~m), Wettzell
(20~m). The data processing technique and imaging results are described by
\cite{PK07}.

This long-term RDV program is one of the best choices for a
large project to measure two-frequency core shifts 
{\referee
on the basis of open archival raw VLBI data
}
for several
reasons: (i) it is optimized to have a good ($u$,$v$)-coverage, (ii) it
has the maximum possible resolution for ground-based VLBI at these
frequencies, (iii) the frequency ratio between the simultaneously
observed bands is high (3.7), and (iv) the core shift per unit of
frequency between 2.3 and 8.6~GHz is larger than that at higher
frequencies because of opacity effects \citep[see, e.g.,][]{L98}.

We have measured the frequency-dependent core shift by model-fitting the
source structure with two-dimensional Gaussian components
\citep{pearson1999} and referencing the position of the core component
to one or more jet features, assuming the latter to be optically thin
and having frequency-independent peak positions (Fig.~\ref{f:art}). The
method is illustrated in Fig.~\ref{f:J1642}, where astrometric S/X
images of \object{1642+690} are shown, with positions marked for the
VLBI core (`C') and jet feature (`A').

\begin{figure}[b!]
\begin{center}
\resizebox{1.0\hsize}{!}{
   \includegraphics[trim=0cm 0cm 0cm 0cm]{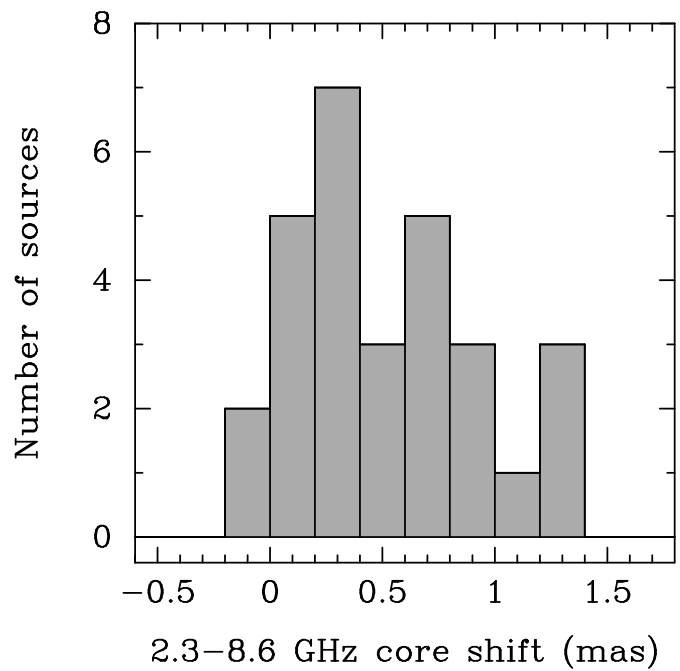}
}
\end{center}
\caption{
\label{f:hist_shifts}
Histogram of the derived core shift values for 29 sources.
One average core shift value per source is used.
The median value for the distribution is equal to 0.44~mas.
}
\end{figure}

\begin{table*}
\caption{Results of core shifts measurements for 29 bright extragalactic
radio sources}
\label{t:shifts}
\begin{center}
\renewcommand{\footnoterule}{}
\begin{tabular}{lccrccc}
\hline
\hline
IAU name   & J2000.0  & J2000 & redshift & Observing & Core shift$^\mathrm{a}$ (mas) \\
           & R.A.     & Decl. &          & epoch     & between 2.3 and 8.6 GHz \\
\hline
\object{0003$-$066}              & 00:06:13.893 & $-$06:23:35.335 &   0.347 & 2003-05-07 & $+0.375 \pm 0.019$ \\
                                 &              &                 &         & 2003-09-17 & $+0.246 \pm 0.028$ \\
\object{0118$-$272}              & 01:20:31.663 & $-$27:01:24.652 &$>$0.557 & 2003-03-12 & $+0.791 \pm 0.075$ \\
\object{0148$+$274}              & 01:51:27.146 & $+$27:44:41.794 &   1.26  & 2002-09-25 & $+1.229 \pm 0.018$ \\
\object{0202$+$149}              & 02:04:50.414 & $+$15:14:11.044 &   0.405 & 2002-03-06 & $+0.058 \pm 0.043$ \\
\object{0239$+$108}              & 02:42:29.171 & $+$11:01:00.728 &   2.680 & 2003-06-18 & $+0.868 \pm 0.032$ \\
\object{0342$+$147}              & 03:45:06.417 & $+$14:53:49.558 &   1.556 & 2002-01-16 & $+0.274 \pm 0.034$ \\
\object{0425$+$048}              & 04:27:47.571 & $+$04:57:08.326 &   0.517 & 2002-09-25 & $+0.385 \pm 0.046$ \\
\object{0430$+$052}              & 04:33:11.096 & $+$05:21:15.619 &   0.033 & 2003-06-18 & $+1.196 \pm 0.018$ \\
                                 &              &                 &         & 2003-09-17 & $+0.956 \pm 0.016$ \\
\object{0507$+$179}              & 05:10:02.369 & $+$18:00:41.582 &   0.416 & 2002-07-24 & $+0.383 \pm 0.012$ \\
\object{0607$-$157}              & 06:09:40.950 & $-$15:42:40.673 &   0.324 & 2002-01-16 & $+0.140 \pm 0.038$ \\
                                 &              &                 &         & 2002-01-16 & $+0.149 \pm 0.065$ \\
\object{0610$+$260}              & 06:13:50.139 & $+$26:04:36.720 &   0.580 & 2002-12-11 & $+1.362 \pm 0.030$ \\
                                 &              &                 &         & 2002-12-11 & $+1.370 \pm 0.030$ \\
\object{0736$+$017}              & 07:39:18.034 & $+$01:37:04.618 &   0.191 & 2003-05-07 & $+0.186 \pm 0.083$ \\
\object{0839$+$187}              & 08:42:05.094 & $+$18:35:40.991 &   1.272 & 2003-03-12 & $+0.911 \pm 0.017$ \\
\object{0850$+$581}$^\mathrm{b}$ & 08:54:41.996 & $+$57:57:29.939 &   1.322 & 1997-01-10 & $+1.349 \pm 0.031$ \\
\object{0952$+$179}              & 09:54:56.824 & $+$17:43:31.222 &   1.478 & 2002-01-16 & $+0.725 \pm 0.039$ \\
\object{1004$+$141}              & 10:07:41.498 & $+$13:56:29.601 &   2.707 & 2002-01-16 & $+0.399 \pm 0.054$ \\
                                 &              &                 &         & 2003-05-07 & $+0.490 \pm 0.034$ \\
\object{1049$+$215}              & 10:51:48.789 & $+$21:19:52.314 &   1.300 & 2002-01-16 & $+0.997 \pm 0.068$ \\
\object{1147$+$245}              & 11:50:19.212 & $+$24:17:53.835 &   0.200 & 2003-05-07 & $+0.321 \pm 0.017$ \\
\object{1219$+$285}              & 12:21:31.691 & $+$28:13:58.500 &   0.102 & 2002-03-06 & $+0.239 \pm 0.008$ \\
\object{1642$+$690}              & 16:42:07.849 & $+$68:56:39.756 &   0.751 & 2002-01-16 & $+0.475 \pm 0.024$ \\
                                 &              &                 &         & 2002-03-06 & $+0.437 \pm 0.025$ \\
                                 &              &                 &         & 2003-05-07 & $+0.350 \pm 0.039$ \\
\object{1656$+$053}              & 16:58:33.447 & $+$05:15:16.444 &   0.879 & 2002-09-25 & $-0.010 \pm 0.048$ \\
\object{1655$+$077}              & 16:58:09.011 & $+$07:41:27.541 &   0.621 & 2002-07-24 & $+0.672 \pm 0.049$ \\
\object{1803$+$784}              & 18:00:45.684 & $+$78:28:04.018 &   0.680 & 2002-05-08 & $+0.722 \pm 0.022$ \\
                                 &              &                 &         & 2002-05-08 & $+0.645 \pm 0.053$ \\
\object{1830$+$285}              & 18:32:50.186 & $+$28:33:35.955 &   0.594 & 2002-03-06 & $+0.772 \pm 0.012$ \\
\object{1845$+$797}              & 18:42:08.990 & $+$79:46:17.128 &   0.057 & 2002-01-16 & $+0.551 \pm 0.049$ \\
                                 &              &                 &         & 2002-01-16 & $+0.403 \pm 0.083$ \\
\object{1926$+$087}              & 19:28:40.855 & $+$08:48:48.413 &   \dots & 2002-03-06 & $-0.088 \pm 0.024$ \\
\object{2021$+$317}              & 20:23:19.017 & $+$31:53:02.306 &   \dots & 2002-07-24 & $+0.158 \pm 0.105$ \\
\object{2143$-$156}              & 21:46:22.979 & $-$15:25:43.886 &   0.698 & 2002-05-08 & $+0.121 \pm 0.018$ \\
\object{2155$-$152}              & 21:58:06.282 & $-$15:01:09.328 &   0.672 & 2003-09-17 & $+0.369 \pm 0.018$ \\
\hline
\end{tabular}
\end{center}
{\bf Column designation:}
Col.~1~-- IAU source name (B1950),
Col.~2~-- Right Ascension (J2000),
Col.~3~-- Declination (J2000),
Col.~4~-- redshift as given by \cite{VV12} with the following exceptions:
the redshifts for \object{0239+108}, \object{0342+147}, \object{1147+245},
\object{2320+506} are taken from \cite{SE_etal05}; the redshift for
\object{0425+048} is taken from \cite{Afanas_etal03},
Col.~5~--- epoch at which the shift was measured (YYYY-MM-DD),
Col.~6~-- shift of the VLBI core position (mas) between 2.3 and 8.6 GHz
and its $1\sigma$ error.
If the shift is measured relative to two different jet features, two
independent values for the same epoch are given.
\\
$^{\mathrm{a}}$ 
Positive values of the shift mean that the distance between the core
and the jet feature measured at 8.6~GHz is greater than the one measured
at 2.3~GHz.
\\
{\referee
$^{\mathrm{b}}$ The measurement for this source is based 
on our imaging results of the NRAO archival data from the VLBA only
S/X project BF025 \citep[see description of the BF025 program in][]{FC00}.
}
\end{table*}

\begin{figure}[t!]
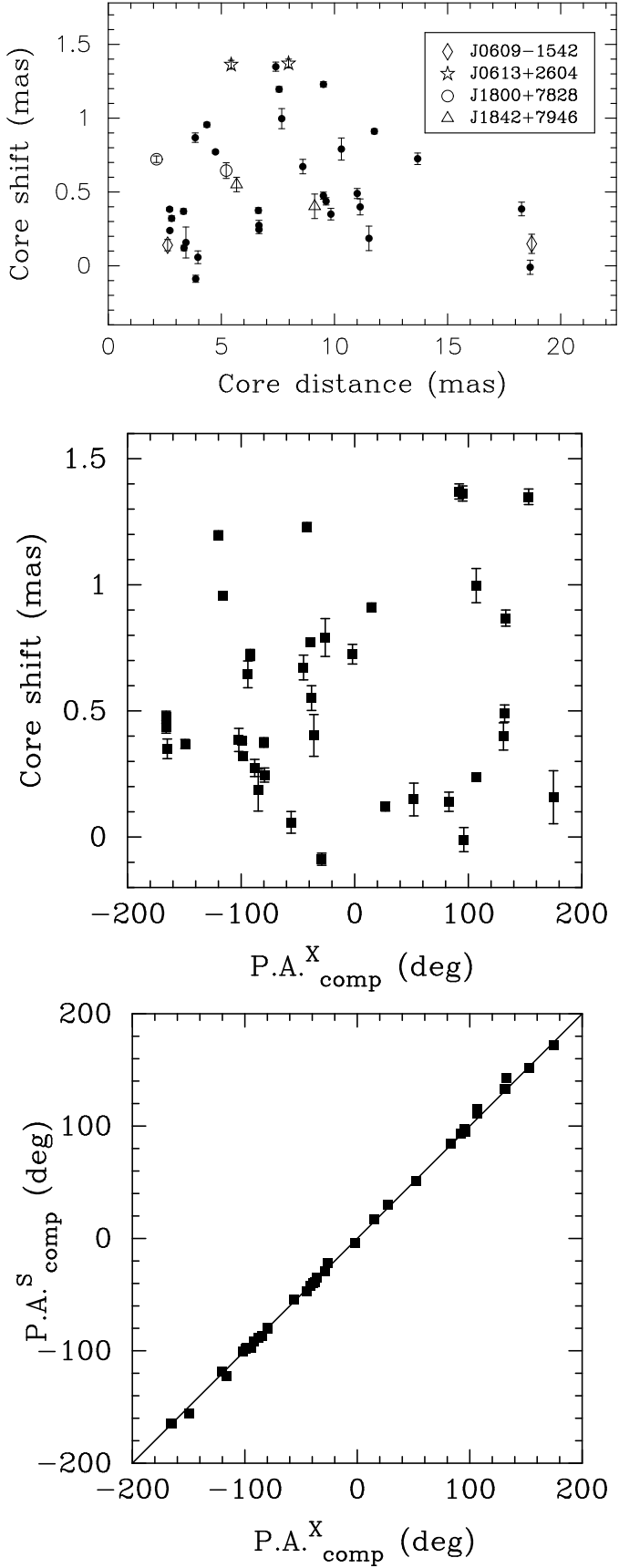

\begin{center}
\resizebox{\hsize}{!}{
   \includegraphics[trim=0cm 0cm 0cm 0cm,angle=-90]{csvsdist_trust.eps}
}
\resizebox{0.98\hsize}{!}{
   \includegraphics[trim=0cm 0cm 0cm -0.5cm]{csvsPA.eps}
}
\resizebox{0.98\hsize}{!}{
   \includegraphics[trim=0cm 0.2cm 0cm -0.1cm]{PAvsPA.eps}
}
\end{center}
\caption{
\label{f:par_deps}
Dependencies between several derived parameters to check for
possible systematics presented in results.
}
\end{figure}

D.~C.~Homan and Y.~Y.~Kovalev (in prep.) have made tests comparing core
shifts measured in the quasars \object{1655$+$077} and
\object{2201$+$315} with relative astrometry (phase referencing to a
calibrator source) to those obtained as here from self-calibrated images by
referencing the core to optically thin jet features. The core shifts
obtained using these two methods agreed within the errors. This result
indicates that self-calibrated images provide sufficient accuracy for
measuring the core shift and can be used for a systematic study of this
effect. However, the VLBI phase referencing method is required for jets
whose structure does not allow the self-referencing technique to be
applied.

Other approaches have also been tried. \cite{Rioja_etal05} have proposed
a ``source/frequency phase referencing'' method for measuring the core
shift, which could be particularly useful while working at higher radio
frequencies. \cite{Walker_etal00} performed a 2-D cross-correlation
analysis in order to align VLBA images of \object{3C~84} measured at
different frequencies with a formal accuracy less than
5~$\mu$as, see also discussion of this method by \cite{CG08}.
However, the real error margins of the cross-correlation method
are certainly substantially larger, as this method requires the
assumption that there are no large variations in observed spectral index
along the jet (if the core region is blended) which is not the case for
many extragalactic jets (see, e.g., the reconstructed spectral index
gradient along the jet for \object{3C~120} in Fig.~\ref{f:3C120spi}).

\begin{figure}[]
\begin{center}
\resizebox{1.0\hsize}{!}{
   \includegraphics[trim=0cm 0cm 0cm 0cm]{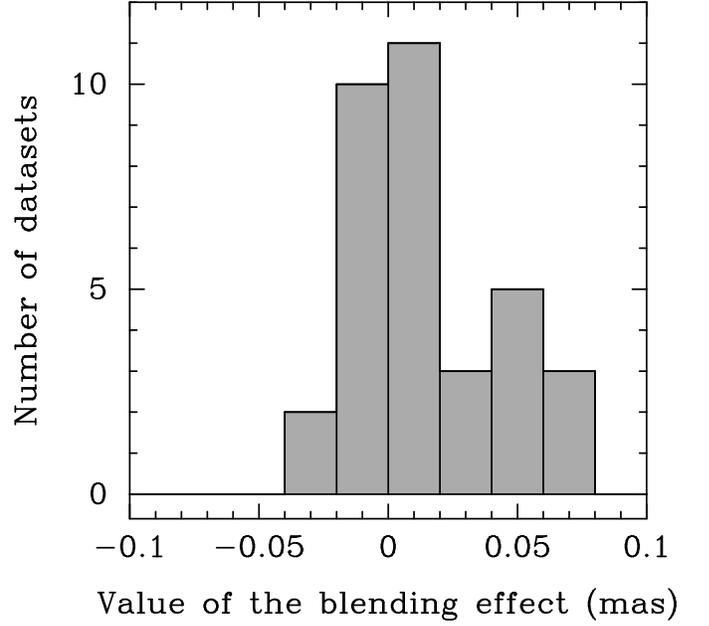}
}
\end{center}
\caption{
\label{f:blendXtest}
Distribution of blending shifts of the core position. The distribution
is calculated for all the sources and epochs in Table~\ref{t:shifts} at
8.6~GHz by downgrading the resolution to the one at 2.3~GHz. Positive
values correspond to inward shifts of the core.
}
\end{figure}

We have measured core shifts between 2.3 and 8.6 GHz in 29 AGN (see
Fig.~\ref{f:hist_shifts} and Table~\ref{t:shifts}), with the resulting
values of the shift ranging between $-0.1$ and 1.4~mas and the median value
for the sample of 0.44~mas. Errors of the core shift measurements
presented in Table~\ref{t:shifts} are estimated from the uncertainties
of component positions calculated following \cite{F99}.
{\referee
The robustness of the measurements is supported by the following
observation. In objects with several bright jet components, similar
core shift values within errors  are obtained from referencing the core
position to different jet features.
The core shift measurements for these objects are indicated by separate 
symbols in the top panel of Fig.~\ref{f:par_deps}.
In objects observed more than once, comparable core shift
values are obtained from referencing the core position to the same jet
feature at different observing epochs (see sect.~\ref{s:variability} for
details on a possible core~shift variability).
In both cases, no systematic dependencies are observed
(Fig.~\ref{f:par_deps}), implying that 
measured core~shift values do not depend on the component location and
features cross-identified between the two frequency bands S/X 
have similar position angles ($\mathrm{P.A.}^\mathrm{S/X}_\mathrm{comp}$).
}
In order to further ensure the fidelity of the measurements made, we
have also performed an independent check by analyzing spectral index
images with and without core shift applied and dropping all objects in
which application of the core shift resulted in unphysical values of
the spectral index
{\referee
(see details below).
}

The difference in resolution and corresponding angular size of the
restoring VLBI beam at 2.3~GHz and 8.6~GHz could lead to a systematical
apparent (not~real) shift of the core position due to
resolution-dependent blending of the core and optically thin parts of
the jet. To estimate the magnitude of this effect, we have downgraded
the resolution of 8.6~GHz data to the one of the 2.3~GHz band by setting
the maximum ($u$,$v$)-radius at 8.6~GHz to about 75~M$\lambda$. Then the
8.6~GHz model was re-fitted for every source and epoch from
Table~\ref{t:shifts} by varying all free parameters. These models were
compared to the original 8.6~GHz ones, and the magnitude of the core
shift due to blending was calculated. Distribution of these blending
shifts of the core position along the jet axis at 8.6 GHz is presented in
Figure~\ref{f:blendXtest}. The median value for this blending effect is
$\approx 0.006$~mas which is much less than the typical error of the
core-shift measurements presented in this paper (Table~\ref{t:shifts}).

{\referee
For about 90~percent of the 277 objects imaged, no reliable estimates of core
shifts have been obtained. All of these objects satisfy one or more of the
following five conditions:

\begin{enumerate}
\item[1)] The core and jet features could not be extracted reliably at one or
both frequency bands because of components blending, or a smooth or weak
jet.

\item[2)] The same structures at both bands could not be identified reliably.

\item[3)] Modeling by two-dimensional Gaussians is not adequate due to the
complexity of the observed brightness distribution.\\
Decision on every of the items 1) to 3) is made on the basis of our analysis
of two-frequency model parameters and their errors,
$\chi^2$ values of fits of models to calibrated visibility datasets,
as well as comparison of residual images after a model is subtracted
with the imaging noise \citep{F99,pearson1999}.

\item[4)] Jet components may be partially opaque 
(spectral index $\alpha\gtrsim0$, $S\propto\nu^\alpha$)
and as a result of this their positions change with frequency.

\item[5)] Application of derived core shift value results in unphysical values
of the spectral index. This implies large optical depth observed in
extended jet regions ($\alpha\gtrsim0$), and/or optically
thin core region ($\alpha\lesssim0$), and/or steep and irregular
$\alpha$ gradients across the jet.
\end{enumerate}

\noindent
The goal of this study is to
provide and analyze most reliable core shift detections and therefore
such stringent selection criteria are applied to reduce the chance of
an erroneous measurement. Based on these considerations, we deem
acceptable the relatively low percentage (about 10~percent of the
sample) of positive measurements. A designated study with
significantly higher rate of successful core shift detections would
require dedicated, time consuming, multi-band VLBI experiments with
phase referencing. We estimate that about 6~hours of VLBA time per
source will be needed in this case, assuming the current sustainable
recording data rate of 128~Mbps. The time requirements would be
shorten significantly after the VLBA recording system is upgraded for
operations at a 4~Gbps recording rate.}

\section{Physics and astrometry with the core shift \label{s:pa}}

The effect of frequency-dependence of the core position has an
immediate connection to several physical and astrometric studies using
compact extragalactic radio sources.  Systematic observations of the
core shift in a sample of compact jets can be used for probing the
conditions in the compact jets and nuclear regions in AGN and
understanding the effect the shift may have on the alignment of the
radio and optical reference frames.

\subsection{Deriving physical properties of the opaque jet base
and absorbing material \label{s:phys}}

If the core shifts and the power index $k_\mathrm{r}$ are measured
from VLBI observations at more than two frequencies, the
magnetic field strength and distribution can be reconstructed in the
ultra-compact regions of the jets \citep{L98}. The offset of the
observed core positions from the true base of the jet as well as the
distance from the nucleus to the jet origin can also be derived \citep{L98}. 
Estimates of the total (kinetic + magnetic field) power, synchrotron
luminosity and the maximum brightness temperature, $T_\mathrm{b,max}$,
in the jets can be made. The ratio of particle energy to magnetic field
energy can also be estimated, from the derived $T_\mathrm{b,max}$. This
information enables testing of the \cite{Koenigl81} model and several of
its later modifications \citep[e.g.,][]{HM86,BM96}. The estimated
distance from the nucleus to the jet origin constrains the
self-similar jet model \citep{Marscher95} and the particle-cascade model
\citep{BL95}. This approach can also be applied to determine the
matter content in parsec-scale jets \citep{hirotani2000,hirotani2005}.

If the core shifts are measured at four or more frequencies, the
absolute geometry of the jet can be determined, giving the absolute
offset, $r_\mathrm{core}$ [pc], of the core from the central engine. Combined
with the value for magnetic field, $B_\mathrm{core}$ [G], this gives an
estimate for the mass of the central black hole $M_\mathrm{bh}{\rm
[M_\odot]} \approx 7\times 10^9 B_\mathrm{core}^{1/2}
r_\mathrm{core}^{3/2}$ \citep{L98}.

Density and pressure gradients in the jet and external absorption can
lead to deviations in $k_\mathrm{r}$ from unity. If the changes in
$k_\mathrm{r}$ with frequency are measured with sufficient precision,
they can be used to estimate the size, particle density and
temperature of the absorbing material surrounding the jets
\citep{L98}. These estimates can be compared with the black hole
masses and BLR sizes obtained from reverberation mapping and
applications of the relation established between the black hole mass
and various kinematic properties of central regions in AGN
\citep[c.f.,][]{PB06}.

\subsection{Nuclear flares and core shift\label{s:variability}}

\begin{figure}[b!]
\begin{center}
\resizebox{0.9\hsize}{!}{
   \includegraphics[trim=0cm 0cm 0cm -0.1cm,angle=0]{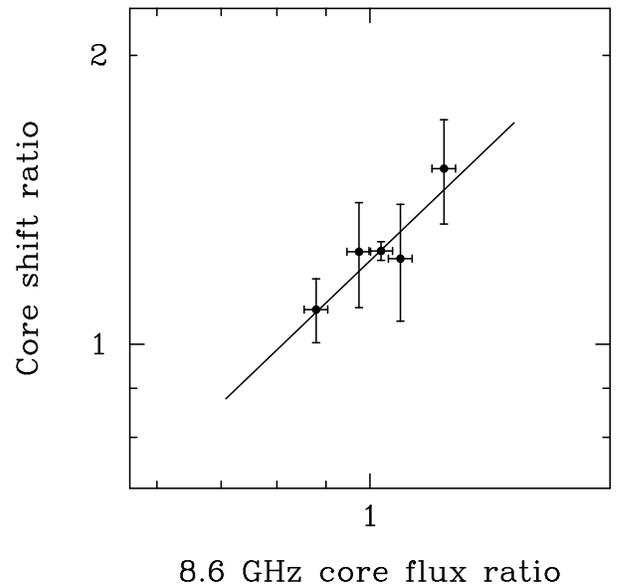}
}
\end{center}
\caption{
\label{f:ratioX}
\referee Core shift ratio versus 8.6~GHz core flux density ratio. The
ratios are calculated for the four sources in the sample which have
measurements at more than one observing epoch (Table~\ref{t:shifts}).
The core shift ratio is defined to be greater than unity, {\em i.e.},
for each pair of measurements, the larger value of the shift is in the
numerator of the ratio.  The line is a least square fit to the
points, yielding a slope of $0.96\pm0.18$.  }
\end{figure}

Opacity in the compact jet can be changed substantially if the source
undergoes a nuclear flare. This can affect the observed position of
the core at a given frequency and modify the core shift between
different frequencies. The magnitude of this effect can be estimated
using the following argument.  Let us assume that, during a flare, the
compact jet does not change its orientation and opening angle. Then,
following \cite{L98}, the magnitude of the core shift is
\begin{equation}
\Delta r \propto S_\mathrm{core}^{2/3} \delta_\mathrm{j}^{2/3}
\beta_\mathrm{j}^{-2/3} \Gamma_\mathrm{j}^{-4/3} \,,
\end{equation}
where $S_\mathrm{core}$ is the flux density
of the core and $\beta_\mathrm{j}$, $\Gamma_\mathrm{j}$,
$\delta_\mathrm{j}$ are the speed, Lorentz factor and Doppler factor
of the jet, respectively.  The flux density of the core is
$S_\mathrm{core} \propto \delta_\mathrm{j}^{2-\alpha} N_0
B_\mathrm{core}^{1-\alpha}$, where $B_\mathrm{core}$ is the magnetic
field in the core region, $N_0$ is the particle density of the jet
plasma and $\alpha$ is the spectral index. It follows immediately that
\begin{equation}
\Delta r \propto (\delta_\mathrm{j}^{3-\alpha}\,N_0\,
B_\mathrm{core}^{1-\alpha}\, \beta_\mathrm{j}^{-1}\,
\Gamma_\mathrm{j}^{-2})^{2/3}\,.
\end{equation}
It has been shown that nuclear flares
are likely to be produced by particle density variations in the
compact jet, while the Doppler factor and magnetic field change only
weakly \citep{Lobanov1999}. Thus, it can be expected that, in the
simplest case $\Delta r \propto N_0^{2/3} \propto
S_\mathrm{core}^{2/3}$. This proportionality can be measured from
comparison of core shifts measured in the same object at different
epochs, yielding a power index $\epsilon = \log(\Delta r_1/\Delta
r_2)/\log(S_1/S_2)$.

\begin{figure*}[t!]
\begin{center}
\resizebox{\hsize}{!}{
   \includegraphics[trim=0cm 4.0cm 0cm 5cm,clip]{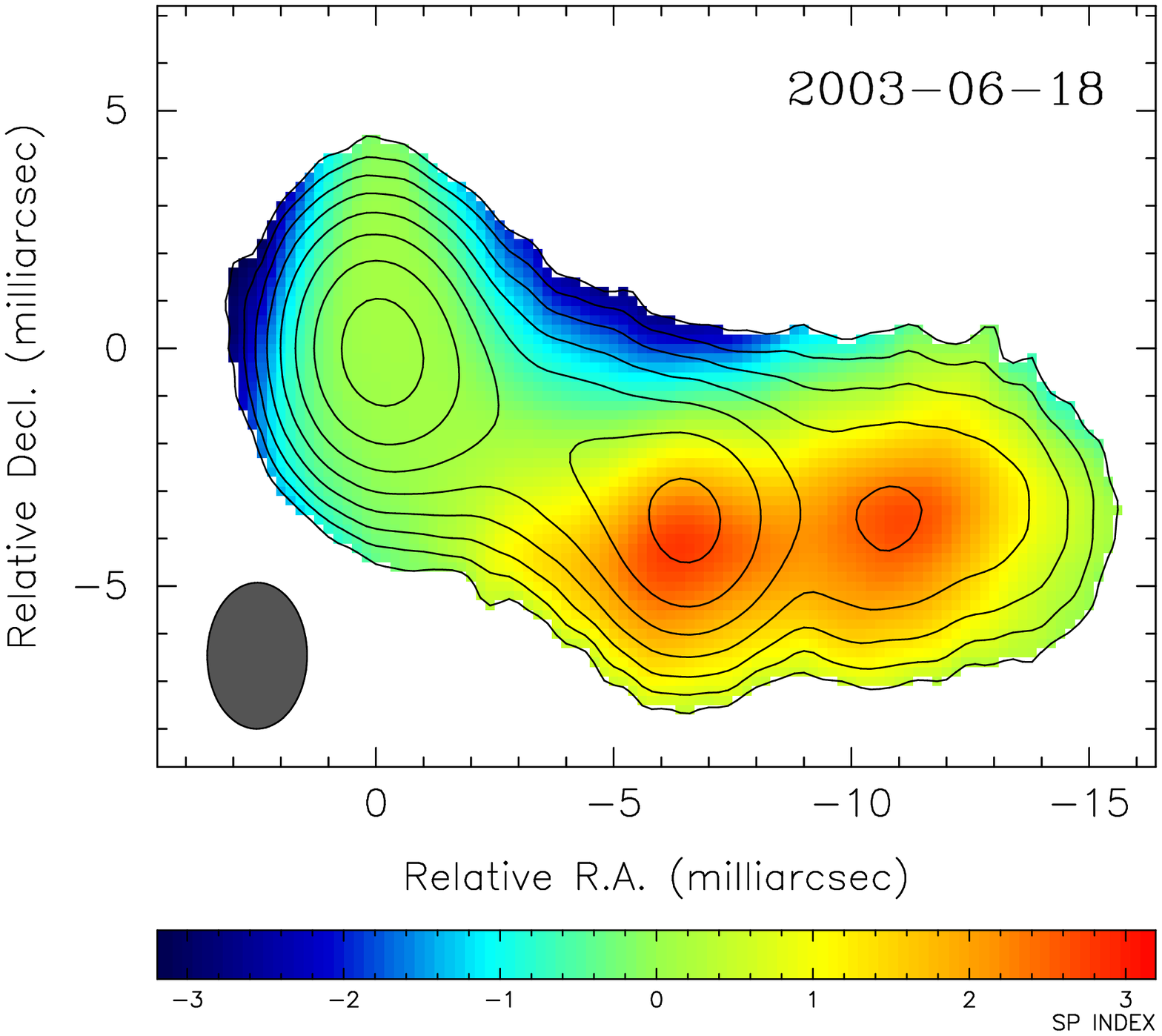}
   \includegraphics[trim=0cm 4.0cm 0cm 5cm,clip]{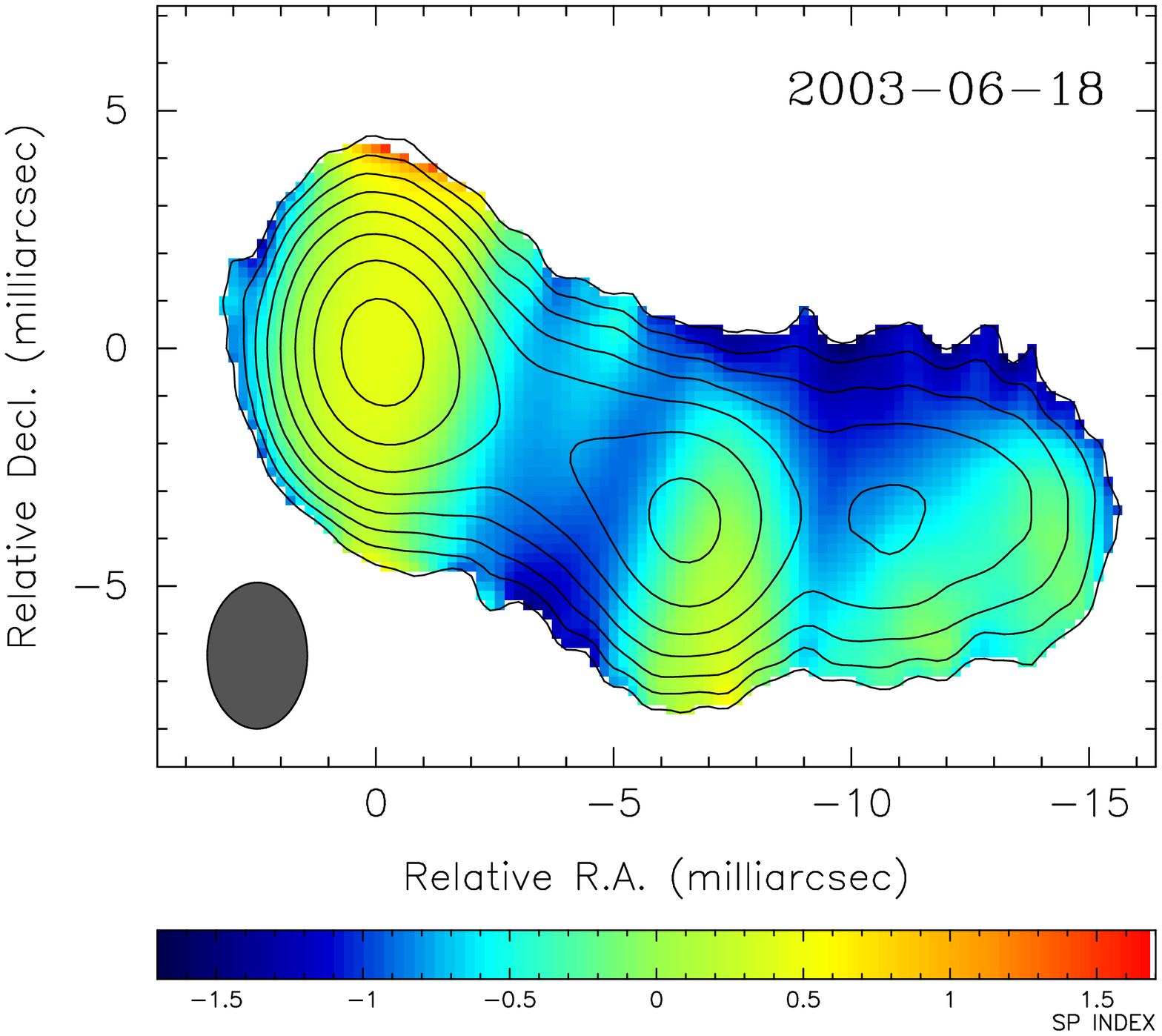}
}
\resizebox{0.5\hsize}{!}{
   \includegraphics[trim=0cm 4.8cm 0cm 5cm,clip]{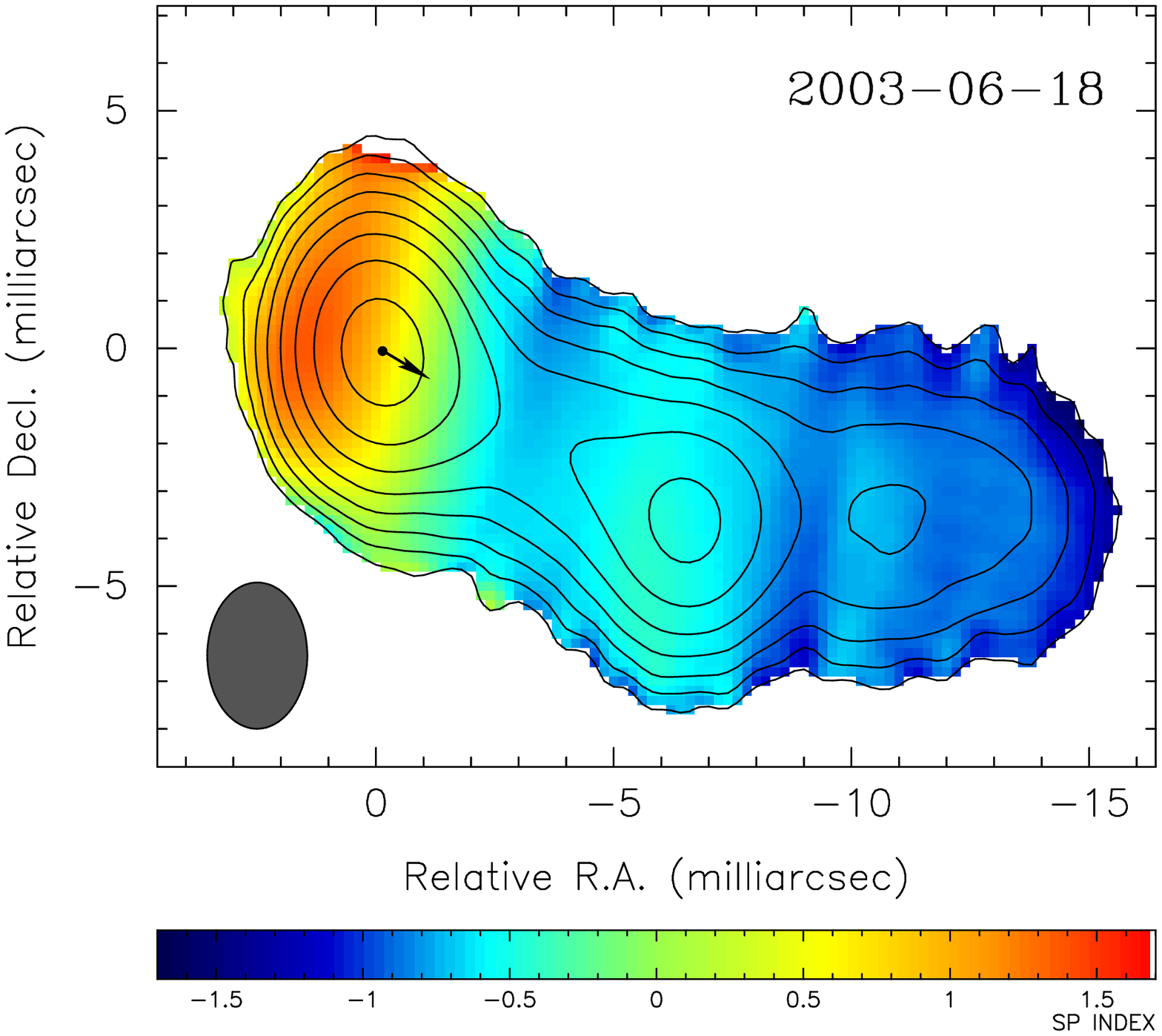}
}
\end{center}
\caption{\label{f:3C120spi}
Influence of the core shift on positional alignment of multi-frequency
VLBI images, illustrated by 2--8~GHz spectral index, $\alpha$
($S\propto\nu^{\alpha}$), images for the jet in \object{3C~120}. Global
VLBI observations at 2.3 and 8.6~GHz made on 18~June~2003, were used.
In each panel, contours represent the total intensity distribution at
8.6~GHz. The restoring beam used for convolution of the total intensity
images at both frequencies is shown in the lower left hand corner of
each map. The spectral index image on the top left is made by aligning
the 2.3 and 8.6~GHz images at their respective phase centers.  The top
right image is aligned at model-fitted core positions. The bottom image
is produced by aligning optically thin jet components (the arrow marks
the resulting core shift between the 2.3 and 8.6~GHz images). 
}
\end{figure*}

In our data, core shifts can be compared at multiple epochs only in
four objects, at five instances altogether (see
Table~\ref{t:shifts}). The ratios of the core shifts measured at
different epochs in the same objects are plotted in
Fig.~\ref{f:ratioX} against the respective ratios of the core flux
density at 8.6~GHz. {\referee Ideally, the ratios of core shifts
should approach unity as the flux density ratio approaches unity. This
is not the case in Fig.~\ref{f:ratioX}, implying that flares in
compact jets involve changes not only in the density of the plasma but
also in its physical properties. This can be further examined by
analyzing the slope of the relation plotted in Fig.~\ref{f:ratioX}.}

The measurements plotted in Fig.~\ref{f:ratioX} indicate
a positive trend, with larger core shifts for higher core flux density. 
A linear fit shown in Fig.~\ref{f:ratioX} corresponds to $\epsilon =
0.96\pm0.18$, which is larger than the value of 2/3 expected for nuclear
flares resulting entirely from plasma density variations in the flow.
This may imply that flaring jets contain a hotter plasma or stronger
magnetic field. However, the small number of data points in the present
plot results in large errors and precludes making any conclusions about
the physical meaning of the fit, and we present it here only as an
example of the effect nuclear flares may have on the core shift. This
effect certainly deserves a more comprehensive study.

\subsection{Effect of the core shift on multi-frequency VLBI studies \label{s:multiVLBI}}

\begin{figure}[tb]
\begin{center}
\resizebox{\hsize}{!}{
   \includegraphics[trim=2cm 0.5cm 0cm 1.5cm,angle=270,clip]{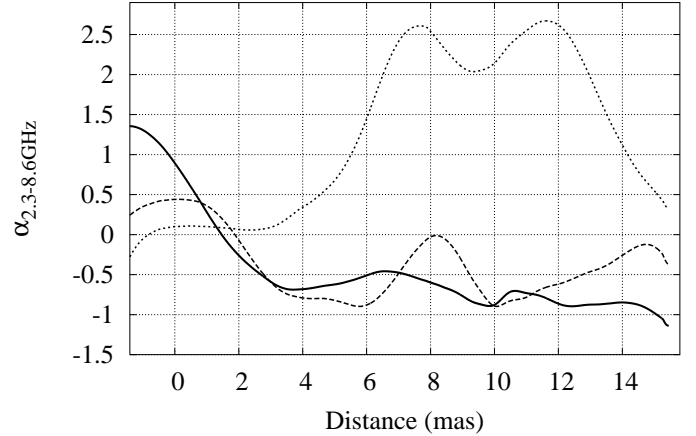}
}
\end{center}
\caption{
\label{f:3C120spi_ridge}
Spectral index profiles along the jet ridge line obtained from the
three different image alignments illustrated by the spectral index
images of \object{3C~120} presented in Fig.~\ref{f:3C120spi}. The
dotted line corresponds to the alignment at the phase
centers. The dashed line corresponds to the alignment at the
model-fitted core positions. The solid line corresponds to the
alignment at the model-fitted positions of optically thin jet
components.
}
\end{figure}

In VLBI observations, application of the self-calibration technique
based on closure phases \citep[e.g.,][]{Jennison58,RW78} results in
a loss of information about absolute positions of target objects. In
this situation, multi-frequency VLBI images are often aligned at the
apparent position of the core at different frequencies (i.e., assuming
implicitly a zero core shift). In the case of a non-zero core shift, this
approach may undermine the results of any study based on multi-frequency
VLBI data (e.g. spectral imaging and Faraday rotation measurements). It
has been shown that accounting for the core shift is critical for
spectral index imaging \citep{L96,L98} and synchrotron turnover
frequency imaging \citep{Lobanov_etal97,L98suppl}. The effect of the
core shift is illustrated by
Fig.~\ref{f:3C120spi}--\ref{f:3C120spi_ridge} for radio spectral
index images of the jet in~\object{3C~120}. The spectral index image
(Fig.~\ref{f:3C120spi}) and the spectral index profile along the jet
ridge line (Fig.~\ref{f:3C120spi_ridge}) obtained without accounting
for the core shift can clearly result in misleading conclusions about
the physical conditions in the jet flow. Therefore, the alignment of
VLBI images should be given proper consideration in all studies
involving multi-frequency measurements.

At present, the S/X band VLBI astrometry programs
\citep[e.g.,][]{icrf98,icrf-ext2-2004,vcs1,vcs2,vcs3,vcs4,vcs5} employ
an implicit assumption that the source positions are the same for both
bands (i.e., all objects are assumed to be point-like, optically thin
radio sources). However, both the core shift and the source structure
should be taken into account in order to achieve a higher positional
accuracy of VLBI astrometry. This issue regarding the source structure
was discussed by \cite{Charlot90}. The likely time variability of the
core shift (due to flaring activity of the nucleus,
Sect.~\ref{s:variability}) may lead to even larger errors in positional
accuracy of repeated astrometric VLBI experiments (e.g., the RDV
project) and must be also taken into account.

\subsection{Radio-optical alignment of astrometric positions\label{s:radio-optics}}

The core shift issue becomes an even more important factor when the
radio reference frame needs to be connected to an optical reference
frame. So far the link is based on the study of optically bright
radio-emitting stars which are seen both by {\em Hipparcos} and VLBI
\citep[see, e.g.,][]{Lestrade_etal95,JV99,Ros2005}. Future accurate alignment
of the frames has to rely on using compact radio sources in distant
quasars \citep[e.g.,][]{Fey_etal2001,Bourda_etal2007}.

Below, we discuss the alignment problem for compact extragalactic
radio sources. We assume that the dominating component in both the radio
and optical bands is the synchrotron self-absorbed compact jet origin
(core). Broad-band modeling of blazar spectral energy distribution
supports this hypothesis \citep[see, e.g., recent review by][]{B07}.
High-resolution VLBI observations of nearby AGN imply that the jet is
formed and emitting in the radio already at distances of $\le 1000$
gravitational radii from the central engine
\citep[e.g.][]{Junor_etal99,Kadler_etal04}. Thus the physical offset
between the jet base and the central nucleus is much smaller than the
positional shift due to opacity in the jet (the latter can be larger
than 1 pc). This implies that the offset between radio and optical
positions of reference quasars will be dominated by the core shift even
if the optical emission comes from the accretion disk around the central
nucleus.

The magnitude of the core shift, $\Delta r$, between two arbitrary frequencies
$\nu_1$ and $\nu_2$ ($\nu_1 > \nu_2$) caused by synchrotron
self-absorption can be predicted for an object with known synchrotron
luminosity, $L_\mathrm{syn}$, of the compact jet \citep{L98}:
\begin{eqnarray}
\label{eq:offset}
\frac{\Delta r}{\mathrm{[mas]}} & = & 4.56\cdot 10^{-21} \frac{1+z}{\Gamma_\mathrm{j}^2 \phi_0}
\left(\frac{\sin\theta_\mathrm{j}}{\beta_\mathrm{j}(1-\beta_\mathrm{j}\,\cos\theta_\mathrm{j})\, \Theta}\right)^{2/3} \times \nonumber \\
& & \times \left(\frac{D_\mathrm{L}}{\mathrm{[pc]}}\right)^{-1}
\left(\frac{L_\mathrm{syn}}{\mathrm{[erg/s]}}\right)^{2/3} \left(\frac{\nu_1\,\nu_2/(\nu_1 - \nu_2)}{\mathrm{[GHz]}}\right)^{-1}\,,
\end{eqnarray}
where $D_\mathrm{L}$ is the luminosity distance to the object,
$\beta_\mathrm{j}$ and $\Gamma_\mathrm{j}$ are the jet speed and
Lorentz factor, $\theta_\mathrm{j}$ is the angle between the jet
velocity vector and the line of sight, $\phi_0$ is the jet opening
angle and $\Theta = \ln(r_\mathrm{max}/r_\mathrm{min})$ describes the
extent, $(r_\mathrm{min},r_\mathrm{max})$, of the compact jet.

If not corrected for, the core shift will introduce an additional
additive error factor in the alignment of the radio and optical
reference frames.  The magnitude of this error can be estimated from the
following considerations. If the two reference frames are aligned using
a sample of AGN with compact radio jets, the expected mean offset of the
source position caused by the core shift can be calculated from
equation~(\ref{eq:offset}). For this calculation, the sample averages
$\langle D_\mathrm{L}\rangle$, $\langle \Gamma_\mathrm{j}\rangle$ and
$\langle L_\mathrm{syn}\rangle$ can be applied.  A conservative estimate
of the jet opening angle $\phi_0 \le 1/\Gamma_\mathrm{j}^2$ can be made,
assuming that the jet has a Mach number $M_\mathrm{j} \sim
\Gamma_\mathrm{j}$. A typical compact jet extent of
$r_\mathrm{max}/r_\mathrm{min} = 100$ \citep{L98} can be adopted for all
objects. It should also be noted that this parameter contributes to the
result of the calculation only logarithmically.

If the sources are selected from a flux density limited sample, the
distribution of their orientations will be affected by the Doppler
boosting \citep{VC94,2cmPaperIII,Cohen_etal07}, with the resulting maximum
viewing angle given by:
\begin{equation}
\theta_\mathrm{lim} = \arccos \left[ \frac{\Gamma_\mathrm{j} -
(T_0/T_\mathrm{lim})^{\varepsilon}}{(\Gamma_\mathrm{j}^2-1)^{1/2}}\right]\,,
\end{equation}
where $T_0$ is the intrinsic brightness temperature of the jet,
$T_\mathrm{lim}$ is the minimum detectable brightness temperature in
the observations, $\varepsilon = 1/(3-\alpha)$ and $\alpha$ is the
spectral index of the radio emission \citep{Lobanov_etal2000}. In the
cores of the VLBI jets, $T_0 \sim 10^{11}$\,K
\citep{Lobanov_etal2000,2cmPaperIV,Homan_etal06}, and typical VLBI snapshot
observations reach $T_\mathrm{lim} \sim 10^8$\,K \citep{Lobanov_etal2000}.

With the assumptions described above, equation~(\ref{eq:offset}) can
be used to calculate the average core shift expected from observations
of a flux density limited sample of extragalactic objects:
\begin{eqnarray}
\label{eq:av_offset}
\frac{\langle\Delta r\rangle}{\mathrm{[mas]}}
& = &
\frac{1.65\cdot 10^{-21}} {\xi_\theta}
\left(\frac{\langle D_\mathrm{M}\rangle}{\mathrm{[pc]}}\right)^{-1}
\left(\frac{\langle L_\mathrm{syn}\rangle}{\mathrm{[erg/s]}}\right)^{2/3} 
\times \nonumber \\ && \times
\left(\frac{\nu_1\,\nu_2/(\nu_1 - \nu_2)}{\mathrm{[GHz]}}\right)^{-1}\,,
\end{eqnarray}
where $D_\mathrm{M} = D_\mathrm{L}/(1+z)$ is the metric (or ``proper
motion'') distance to the object and $\xi_\theta$ evaluates the term
$[\sin\theta_\mathrm{j}/\beta_\mathrm{j}(1-\beta_\mathrm{j}\,\cos\theta_\mathrm{j})]^{2/3}$
at a ``median'' angle of the sample, $\theta_\mathrm{med}$, determined
from the condition $\int_0^{\theta_\mathrm{lim}} \xi_\theta\,
\mathrm{d}\theta = 2\int_0^{\theta_\mathrm{med}} \xi_\theta\,
\mathrm{d}\theta$.

\begin{table}
\caption{\referee{}Theoretical predictions of expected core shifts values}
\label{tb:offsets}
\begin{center}
\referee
\begin{tabular}{rrcccc}
\hline\hline
$\langle \Gamma_\mathrm{j} \rangle$ & $\theta_\mathrm{lim}$ &
$\theta_\mathrm{med}$ & 
$\langle \Delta\,r_\mathrm{2.3\,GHz}^\mathrm{8.6\,GHz} \rangle$ &
$\langle\Delta\,r_\mathrm{8.6\,GHz}^{6000\,\AA} \rangle$  & 
$\langle L_\mathrm{44}^{\star}\rangle$ \\
& [deg] & [deg] &  [mas] & [mas] & [$10^{44}$\,erg/s]\\
\hline
5  &  107.4  & 37.4 & 0.13  & 0.05 & 6.2 \\               
10 &   69.2  & 23.2 & 0.19  & 0.07 & 3.5 \\               
15 &   55.2  & 17.8 & 0.23  & 0.08 & 2.6 \\               
20 &   47.4  & 14.8 & 0.26  & 0.10 & 2.2 \\               
25 &   42.1  & 12.9 & 0.28  & 0.11 & 2.0 \\               
30 &   38.3  & 11.5 & 0.30  & 0.12 & 1.8 \\               
\hline
\end{tabular}
\end{center}
{\bf Column designation:} $\Gamma_\mathrm{j}$~--- jet Lorentz factor;
$\theta_\mathrm{lim}$~--- maximum viewing angle for detection;
$\theta_\mathrm{med}$~--- median viewing angle for a sample with random
orientation; $\Delta\,r_\mathrm{2.3GHz}^\mathrm{8.6GHz}$~---
average shift expected between 2.3\,GHz and 8.6\,GHz;
$\Delta\,r_\mathrm{8.6GHz}^\mathrm{6000\,\AA}$~--- average shift expected
between 8.6\,GHz and 6000\,\AA. All parameters are calculated for a
sample with $\langle L_\mathrm{syn}\rangle = 10^{44}$\,erg/s,
$\langle D_\mathrm{L}\rangle = 3.32$\,Gpc ($z=1$).
{\referee
The last column lists the mean synchrotron luminosities $\langle
L_\mathrm{44}^{\star}\rangle$ required to reconcile
the predicted $\langle \Delta\,r_\mathrm{2.3\,GHz}^\mathrm{8.6\,GHz}\rangle$
with the median value of the core shift (0.44\,mas) measured in
the data presented in this paper.
}
\end{table}

Table~\ref{tb:offsets} gives average core position shifts expected
between 2.3\,GHz and 8.6\,GHz and between 8.6\,GHz and 6000\,\AA\ for
an average synchrotron luminosity of $10^{44}$\,erg\,s$^{-1}$
\citep{L98} and a quasar sample peaking at a redshift $z=1$.  The
shifts are calculated for several different values of $\langle
\Gamma_\mathrm{j}\rangle$ ranging from 5 to 30. {\referee The average
2.3--8.4\,GHz core shift of 0.3\,mas predicted for
$\langle\Gamma_\mathrm{j}\rangle = 30$ is close to the mean value of
0.44\,mas measured from our data. The direct comparison of these two
values is not strictly valid, as our measurements are not done for a
complete sample. We can note however that this discrepancy can be
easily reconciled by allowing for small variations of the model
parameters (i.e. varying the mean luminosity of the sample, as
illustrated in Table~\ref{tb:offsets}). With this assumption, the
maximum measured value of the 2.3--8.4\,GHz core shift can be readily
explained by a one sigma deviation from the mean in a sample of
compact jets with a Gaussian distribution of Lorentz factors
characterized by $\langle\Gamma_\mathrm{j}\rangle = 30$ and
$\sigma_\mathrm{\Gamma} = 10$.  }

It is immediately seen from Table~\ref{tb:offsets}
that the average shifts between the radio and optical bands are
comparable to the positional accuracy of VLBI and {\em{}GAIA}, and
significantly exceed that of {\em{}SIM}. These numbers, albeit not definite
and assumption-dependent, imply that the core shifts should be carefully
investigated, and corrected for, before attempting to make a connection
between the ICRF and the reference frames that will be produced by
{\em{}GAIA} and {\em{}SIM}. 

To achieve this goal, a coordinated program should be established,
aiming at defining a Primary Reference Sample (PRS) of extragalactic
objects to be used for radio-optical reference frame alignment and
investigating the magnitude and variability of the core shift in this
sample. This would require making multi-frequency and multi-epoch VLBI
observations of the PRS in order to measure the core shifts and monitor
their variability for a period of time comparable with the mission
duration of {\em{}GAIA} and {\em{}SIM}. Additional VLBI observations of the sample
would have to be done contemporaneously with the optical measurements.

The accuracy of the determination of the optical position of the core from
radio measurements can be assessed in the following way. Let there be
$N-1$ measurements of core shift between $N$ radio frequencies
($\nu_1,...,\nu_N$). Then the core position at an arbitrary frequency
can be determined from the core offset measure, $\Omega_\mathrm{r\nu}$
\citep[see definition in][]{L98}, provided that
$\Omega_{\mathrm{r\nu}} \approx const$ for
all measurements (this ensures that the accuracy of extrapolation does
not depend on the actual wavelength to which the core position is
extrapolated). The error, $\sigma_\nu$, can be estimated from
\begin{equation}
\sigma_\nu = \frac{\sigma_1}{N-1} \left[N-1 + \sum_{i=2}^{N} \left(
\frac{\nu_1}{\nu_i}\right)^2\right]^{1/2} ,
\end{equation}
where $\sigma_1$ is the positional measurement accuracy at the lowest
frequency, $\nu_1$, in the data set. For typical VLBI measurements
($\mathrm{SNR}_\mathrm{core}\sim 500$, $\mathrm{SNR}_\mathrm{jet} \sim
100$) spanning four frequencies from 8~GHz to 43~GHz, we estimate
$\sigma_\nu \approx 0.05$~mas, with referencing the core position to an
optically thin feature in the jet (measurements at 2~GHz and 5~GHz may
still be required for monitoring $\Omega_\mathrm{r\nu}$ at lower
frequencies and verifying that it is not variable). For relative
astrometry VLBI measurements with phase referencing, we obtain
$\sigma_\nu \approx 0.04$~mas, assuming that measurements are done at
four frequencies between 5~GHz and 22~GHz and respective positional
errors are the typical ones reported by \cite{Fomalont05} for the case
of one phase reference calibrator at a distance of $3^\circ$ from the
target.
These estimates indicate that extrapolating core shifts to optical
wavelengths can be done with sufficient accuracy for astrometric
applications.

If the core shift effect is not corrected for, reaching the accuracy
goal for the radio-optical alignment would require increasing the number
of reference sources by some factor. Using the average radio-optical
core shift in Table~\ref{tb:offsets}, we estimate this factor to be
about 3.5 and 4.5 for the alignment of the ICRF reference frame with the
reference frames of {\em{}GAIA} and {\em{}SIM}, respectively. These
values may be reduced by about 20~percent, if individual radio-optical
position offsets are weighted appropriately with respect to the radio
jet direction \citep[e.g.,][]{2cmPaperIV} along which the core is
expected to shift. But even then it would require a substantial increase
in the number of reference objects required for reaching the sufficient
accuracy of the alignment.

\section{Summary \label{s:sum}}

Measurements of the frequency-dependent shift of the parsec-scale jet
cores in AGN are reported for 29 bright extragalactic radio sources. It
is shown that the shift can be as high as 1.4~mas between 2.3 and
8.6~GHz. If not taken into account, such core shifts could influence and
corrupt both astrophysical (e.g., spectral imaging and Faraday rotation
imaging) and astrometric studies.
It is shown that nuclear flares can result in temporal variability
of the core shift. Analysis of this effect can shed light on the physics
of the flares.
The core shifts are likely to influence the positional accuracy of the
radio reference frame and pose problems for connecting radio and optical
reference frames.  The next steps in this study are to measure core
shifts in a complete flux-density-limited sample of bright extragalactic
radio sources, to estimate physical parameters from the measured shifts
and to look for their temporal variations. We plan to achieve these
goals by combining together further analysis of RDV VLBA S/X datasets,
four-frequency (8.1, 8.4, 12, \& 15 GHz) MOJAVE-2 VLBA
observations\footnote{\sf{}http://www.physics.purdue.edu/astro/MOJAVE/}
in 2006, and dedicated multi-frequency 1.4--15~GHz VLBA observations in
2007 of more than half of the objects in Table~\ref{t:shifts}.

We have estimated from theory an average shift between the radio (4~cm)
and optical (6000~\AA) bands to be approximately 0.1~mas for a
complete sample of radio selected AGN.  The robustness of this
prediction is supported by the fact that the same method gives an
average core shift of 0.2 to 0.3~mas between 2.3 and 8.6~GHz, which
agrees well with the median shift of 0.44~mas reported by us here for a
non-complete sample of 29 AGN.

The estimated radio-optical core shift is comparable to the positional
accuracy of {\em{}GAIA} and significantly exceeds that of {\em{}SIM}. It
implies that the core shift effect should be carefully investigated, and
corrected for, in order to align accurately the radio and optical
positions. We suggest two possible approaches to tie the radio and
optical reference frames together.
1)
In the first approach, multi-frequency VLBI measurements can be used for
calculating the projected optical positions, assuming that the radio and
optical emission regions are both dominated by a spatially compact
component marginally resolved with VLBI and {\em{}SIM} and point-like
for {\em{}GAIA}.
{\referee
The core~shift measurements of the objects are required to be made
quasi-simultaneously with astrometric measurements of their
positions.
}
The discrepancies between the measured optical and radio positions can
then be corrected for the predicted shifts, and the subsequent alignment
of the radio and optical reference frames can be done using standard
procedures. 
2)
A more conservative approach may also
be applied, by employing the VLBI observations to identify, and
include in the Primary Reference Sample, only those quasars in which
no significant core shift has been detected in multi-epoch
experiments. Either of the two approaches should lead to substantial
improvements of the accuracy of the radio-optical position alignment.

\begin{acknowledgements}

     This work is based on the analysis of global VLBI observations including
the VLBA, the raw data for which were provided to us by the open NRAO
archive. The National Radio Astronomy Observatory is a facility of the
National Science Foundation operated under cooperative agreement by
Associated Universities, Inc.
     Y.~Y.~Kovalev is a Research Fellow of the Alexander von Humboldt
Foundation. Y.~Y.~Kovalev was supported in part by the Russian
Foundation for Basic Research (projects 05-02-17377 and 08-02-00545).
     We would like to thank Patrick Charlot, Ed Fomalont, Jos\'e Carlos
Guirado, Jon Marcaide, Leonid Petrov, Richard Porcas, Eduardo Ros as
well as the NASA GSFC VLBI group and the MOJAVE team for fruitful
discussions.
{\referee We thank the anonymous referee for thoughtful reading and
useful comments which helped to improve the manuscript.}
     This research has made use of the NASA/IPAC Extragalactic Database (NED)
which is operated by the Jet Propulsion Laboratory, California Institute
of Technology, under contract with the National Aeronautics and Space
Administration.
     This research has made use of NASA's Astrophysics Data System.

\end{acknowledgements}

\bibliographystyle{aa}
\bibliography{yyk}

\end{document}